\documentclass[a4paper,11pt]{article}
\pdfoutput=1 

\usepackage{jinstpub} 

\title{SiPM Dynamic Range for CEPC Scintillator-based Electromagnetic Calorimeter}

\author[a,b]{Yazhou Niu}
\author[a,b]{Yukun Shi}
\author[,c]{Hang Zhao}
\author[a,b]{Yunlong Zhang}
\author[c]{Manqi Ruan}
\author[a,b,1]{Jianbei Liu,\note{Corresponding author.}}

\affiliation[a]{State Key Laboratory of Particle Detection and Electronics, (Beijing 100049, Hefei 230026), China}
\affiliation[b]{Department of Modern Physics , University of Science and Technology of China, HeFei 230026, China}
\affiliation[c]{Institute of High Energy Physics, Chinese Academy of Sciences (CAS), Beijing 100049, China}

\emailAdd{liujianb@ustc.edu.cn}

\abstract{A high-granularity scintillator calorimeter readout with silicon photomultipliers (SiPMs) is an electromagnetic calorimeter (ECAL) candidate for experiments at the Circular Electron Positron Collider (CEPC). A critical design parameter of this ECAL candidate is the dynamic range of the SiPMs. This study investigates the SiPM dynamic range required for the CEPC scintillator ECAL. A model is developed on the basis of the operation principles of SiPMs to describe the response of an SiPM to light pulses within one recovery period by considering the cross-talk effect, photon detection efficiency, and number of pixels. The response curve of a 10000-pixel SiPM predicted by the model is consistent with the measured curve within $2\%$ for an incident light pulse of up to 12000 photons.
The intrinsic fluctuations of the SiPM response naturally exist in this model, and the correction of the saturation effect in the SiPM response is investigated.
Monte Carlo (MC) simulation shows that the algorithm can restore the response linearity of an SiPM for an incident light pulse in which the number of photons is up to around six times the number of SiPM pixels.
The model and correction program are implemented for full simulation of the $ZH$ production $Z \rightarrow$$\nu$$\nu$, H$\rightarrow$$\gamma$$\gamma$ channel to evaluate the impact of the SiPM dynamic range of the CEPC scintillator ECAL on the reconstructed Higgs boson mass and the sensitivity to the Higgs signal in this channel.
The results show that the CEPC scintillator ECAL equipped with no less than 4000 SiPM pixels and operated with a light yield of 20 photon-electrons per channel for a single minimum ionizing particle can meet the requirements for Higgs boson precision measurement in the di-photon channel at the CEPC.
}

\keywords{CEPC,electronmagnetic calorimeter,SiPM,digitization}

\arxivnumber{} 

\begin{document}
\maketitle
\flushbottom

\section{Introduction}
\label{sec:intro}

\paragraph{}Following the discovery of the Higgs boson in 2012, precise measurements of the Higgs boson properties are essential of high-energy physics in the coming decades. A lepton collider provides a clean environment for precision measurement of Higgs boson properties as well as excellent sensitivity for new physics beyond the stand model. The Circular Electron Positron Collider (CEPC) is a large international scientific project initiated by China and a candidate for a Higgs boson factory. CEPC will operate at a center-of-mass energy of $\sqrt{s} \sim 240 GeV$, producing over a million Higgs bosons and allowing model-independent measurement of Higgs~\cite{1} couplings. To fully exploit the potential of the CEPC physics program, the invariant mass resolution of W, Z, Higgs bosons with hadronic final states must be better than $4\%$, which requires an unprecedented jet energy resolution~\cite{1}.

\begin{figure}[htbp]
\centering 
\includegraphics[width=1\textwidth]{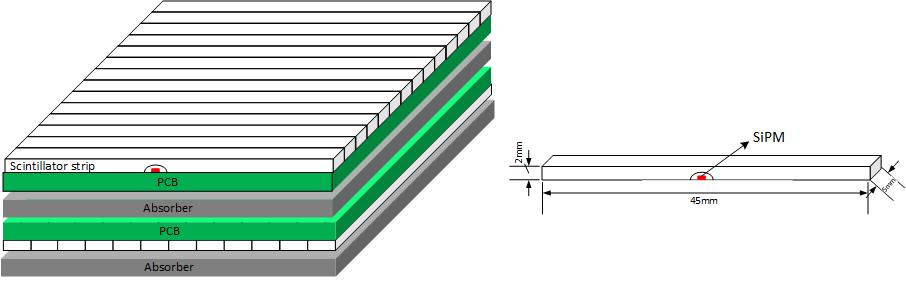}
\caption{\label{fig:layout} Layout of ScECAL and dimensions of scintillator strip coupling with SiPM at bottom-center. Two adjacent sensitive layers are arranged perpendicularly to achieve high lateral granularity.}
\end{figure}

The energy of a typical jet is carried by 65\% charged particles, 25\% photons, and 10\% neutral hadrons. Under the principle of the particle flow algorithm (PFA)~\cite{2,3,4,5,5-1}, each final state particle is reconstructed in the most optimal sub-detector. The momentum of charged particles is determined by the tracking system, while the energies of photons and hadrons are determined by electromagnetic and hadronic calorimeters, respectively.
For a PFA-oriented detector system, the calorimetry system must distinguish individual particles from a jet and determine photons from $0.1 GeV$ to $100 GeV$.
The properties must agree with the PFA principle, and the ECAL is required to have not only good energy resolution but also unprecedented three-dimensional spatial resolution.
Therefore, ScECAL with finely segmented and highly lateral granular features has been designed and optimized by the CALICE collaboration and CEPC calorimeter working group~\cite{6, 7}.
Scintillator strips and silicon photomultipliers (SiPMs) are fabricated as sensitive layers, while tungsten-copper alloy plates as used as absorbers.
The layout of CEPC ScECAl is shown in Figure.~\ref{fig:layout}.

SiPMs, also referred to as multi-pixel photon counters (MPPCs), are novel semiconductor photon detectors. An SiPM is a matrix of pixels (typically, a few 1000 per mm${^2}$); the photodiode operates in the Geiger mode ~\cite{8} so that each pixel works as a photon-independent counter.
In SiPM design, one pixel can avalanche at most once within one recovery period even if multiple photons are incident, which leads to the intrinsically nonlinear saturation response of SiPMs, as has been observed by many groups ~\cite{9,10}. The SiPM response is mainly a function of the number of incident photons and the number of SiPM pixels. The cross-talk and second-time avalanche in the same pixel also affect the SiPM response. The saturation of SiPMs might limit the dynamic range of ScECAL. It is essential to determine the minimum number of pixels required for the CEPC ScECAL. This study focuses on the modeling and validation of the SiPM response and presents a reference SiPM setup for CEPC ScECAL.

We developed a model based on the operation principles of SiPMs to describe the SiPM response by Monte Carlo (MC) simulation.
The real SiPM response curve was measured by an LED-based system and compared with the simulation results of the model over a wide range.
Correction was implemented for the SiPM saturation response to achieve linearity. The uncertainty in the correction was calculated according to the width of the number of fired pixels by MC simulation, which cannot be predicted. Based on the CEPC full simulation and reconstruction tools, the model and correction program were implemented for a full simulation to study the impact of the SiPM response on the benchmark at $ZH$ ($Z \rightarrow$$\nu$$\nu$, H$\rightarrow$$\gamma$$\gamma$).
Finally, a realistic setting was recommended for SiPMs to satisfy the CEPC requirements.

The remainder of this paper is organized as follows. Section~\ref{sec:model} describes the digitization implementation. Section~\ref{sec:valid} presents the real SiPM response measurements results and compares them with the digitization results. Section~\ref{sec:simu} presents the full simulation results of the effect of the SiPM response on the benchmark physics process. Finally, Section~\ref{sec:summary} concludes the paper.

\section{SiPM response and correction}
\label{sec:model}

\paragraph{}Owing to their small size, high gain, and insensitivity to magnetic fields, SiPMs have many potential applications.
However, the saturation response of SiPMs limits their dynamic range to one-third of the total number of pixels. Although increasing the number of pixels can ameliorate the saturation, a dedicated correction is essential to fully exploit the SiPMs pixels.
This section presents SiPM digitization and validation on the basis of the real SiPM response curve measurement results.
In addition, the correction of the SiPM saturation response and the uncertainty estimated in the correction process are discussed.


\subsection{Model of SiPM response}
\label{subsec:digitization}
\begin{figure}[htbp]
\centering 
\includegraphics[width=.6\textwidth]{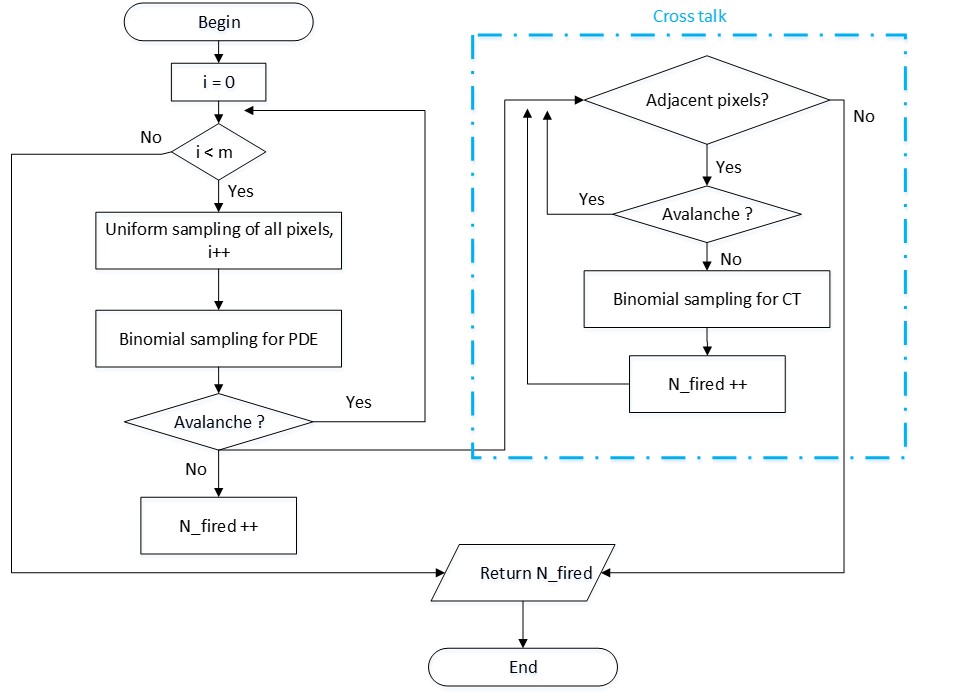}
\qquad
\includegraphics[width=.3\textwidth]{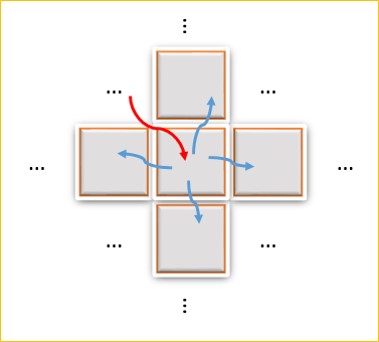}
\caption{\label{fig:logic} \textit{Left}: Implementation of SiPM digital model. \textit{Right}: Schematic diagram of cross-talk.}
\end{figure}

\paragraph{}
The model was developed on the basis of the operation principles of SiPMs and is described as follows. An incident photon is converted into a photon-electron according to the binomial function of the photon detection efficiency. After conversion, the photon-electron will be incident on an arbitrary pixel according to a random sampling function. If the pixel does not receive a photon-electron before that, it will avalanche, increasing the total number of fired pixels by one. One pixel can avalanche at most once within one recovery period even if multiple photon-electrons are incident on one pixel, therefore if the pixel receives a photon-electron before that, nothing will change.
One pixel avalanche has the potential to cause an avalanche in the neighboring pixels. The model will determine if the four adjacent pixels avalanche; if not, a binomial function will be implemented according to the cross-talk probability and the total number of fired pixels will be increased by one if one pixel avalanches.
The logical flow of model is shown in Fig.~\ref{fig:logic}. The program was implemented in C++.
Some initial parameters need to be input to the program, including the incident photons, total number of pixels, photon detection efficiency, and corresponding cross-talk probability.
For a certain number of incident photons, the model will eventually return the total number of fired pixels.
Based on this model, the SiPM response and its intrinsic fluctuation are described well within one recovery time.
If one-pixel discharge completes after the avalanche and a photon-electron is received after one recovery period, an avalanche will occur again, i.e., a second-time avalanche, and it will be added into this model in the next step.
Dark count is another important feature of SiPMs, which is mainly induced by thermally generated electrons. However, the dark count has a trivial effect on the SiPM response when measured with an external trigger; hence, it is not involved in the SiPM response model.

The following equation describes the average number of fired pixels as a function of the number of incident photons~\cite{9, 10}:

\begin{equation}
\label{eq:SiPMformula}
\begin{split}
N_{fired} &= N_{eff}\times[1-e^{\frac{-N_{photon}}{N_{eff}}}]
\end{split}
\end{equation}

where $N_{photon}$ represents the number of effective photons (i.e., incident photons by photon detection efficiency; hereafter, incident photons refers to effective photons), and $N_{eff}$ represents the number of effective pixels, which depends on the number of pixels, cross-talk probability, and second-time avalanche. When all the photons arrive within one recovery period of the SiPM and the cross-talk probability equals 0, $N_{eff}$ is exactly equal to the number of pixels. In this case, the SiPM output is mainly a function of the number of incident photons and number of pixels. In practice, by fitting the real SiPM response curve measurement results, the parameter $N_{eff}$ can be extracted.

\subsection{Validation of SiPM model}
\label{sec:valid}

\begin{figure}[htbp]
\centering 
\includegraphics[width=.45\textwidth]{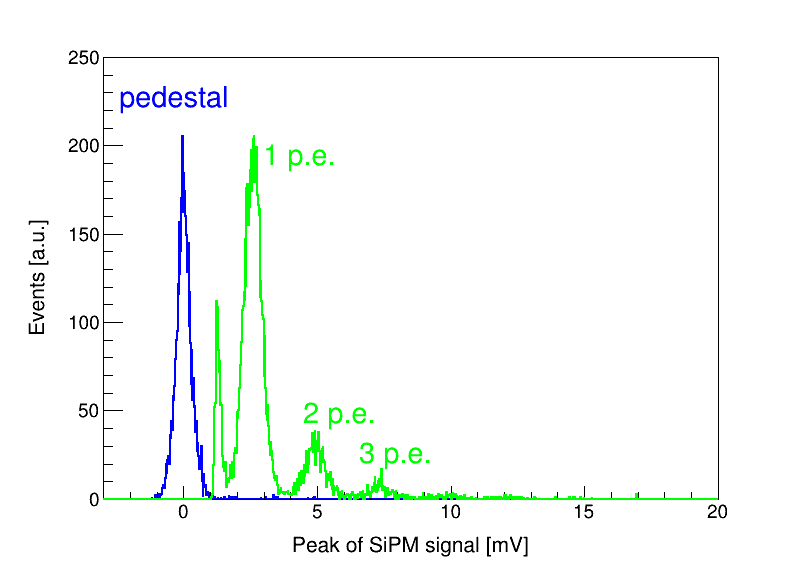}
\qquad
\includegraphics[width=.45\textwidth]{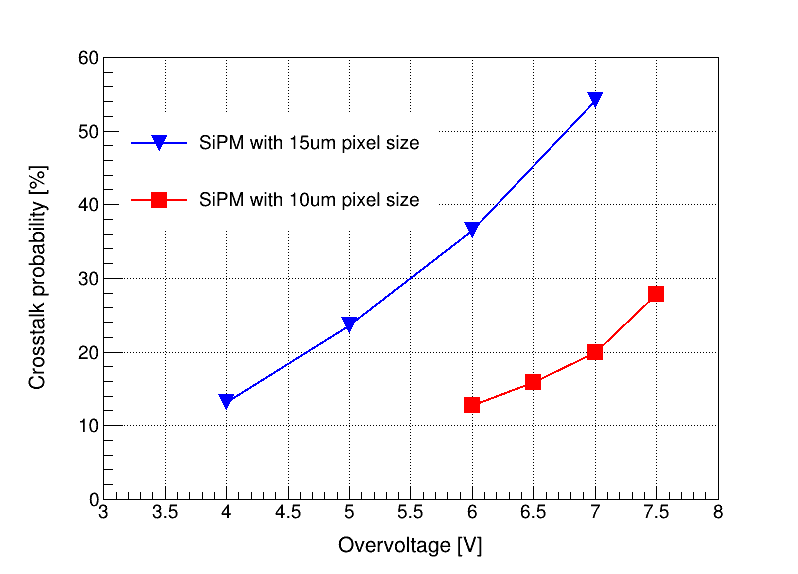}
\caption{\label{fig:SPS_CT} \textit{Left}: Measurement results of single-photon spectrum of SiPM through waveform sampling. \textit{Right}: Cross-talk probability as a function of overvoltage (difference between operation voltage and breakdown voltage).}
\end{figure}

\paragraph{}
Two types of SiPMs produced by HPK~\cite{11}, with 10000 pixels and 4489 pixels, are studied here as examples. For certain SiPMs, the number of pixels is known and the cross-talk probability must be measured.
The cross-talk probability can be determined by measuring the dark count noise of SiPMs.
The two commonly used measurement methods are as follows. The first is to measure the dark count ratio under different discriminator thresholds; the other is to measure the single-photon spectrum.
Here, we use the latter method. We record the waveform of the SiPM signal with a $2.5$-GHz oscilloscope under completely light-shielded conditions. The peak amplitude of each waveform is defined as the SiPM signal value. The 1D histogram distribution of the SiPM signal value is the single-photon spectrum, as shown on the left-hand side of Fig.~\ref{fig:SPS_CT}.
The cross-talk probability is calculated as the ratio of the number of pulses above 1.5 photon-electron (p.e.) and the number of pulses above 0.5 photon-electron. The cross-talk probability increases the SiPM overvoltage, as the gain and photon detection efficiency of the SiPM increases with the SiPM overvoltage. The right-hand side of Fig.~\ref{fig:SPS_CT} shows the cross-talk probability as a function of the overvoltage so that the SiPM should operate at an appropriate voltage. The measurement values of the cross-talk probability are consistent with the datasheet from HPK.

\begin{figure}[htbp]
\centering 
\includegraphics[width=.3\textwidth]{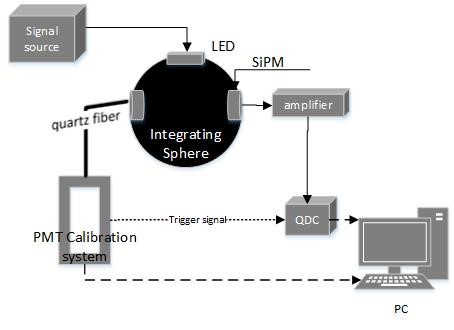}
\quad
\includegraphics[width=.3\textwidth]{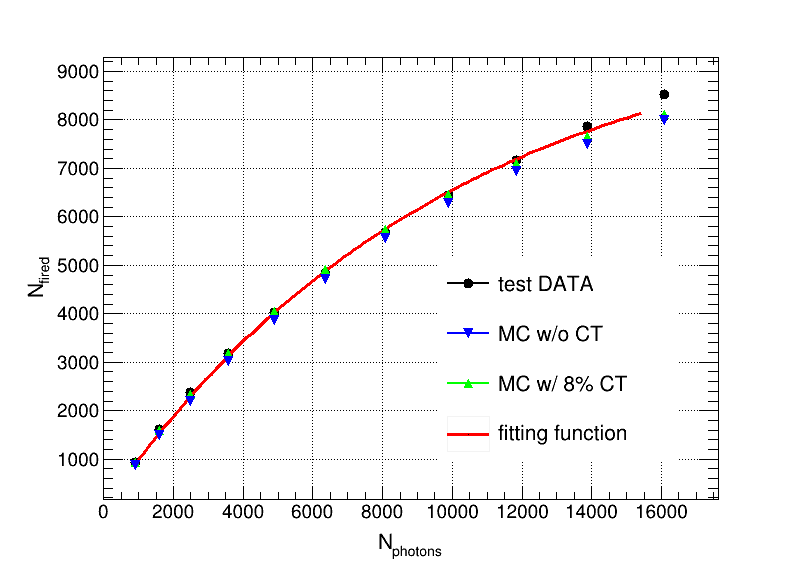}
\quad
\includegraphics[width=.3\textwidth]{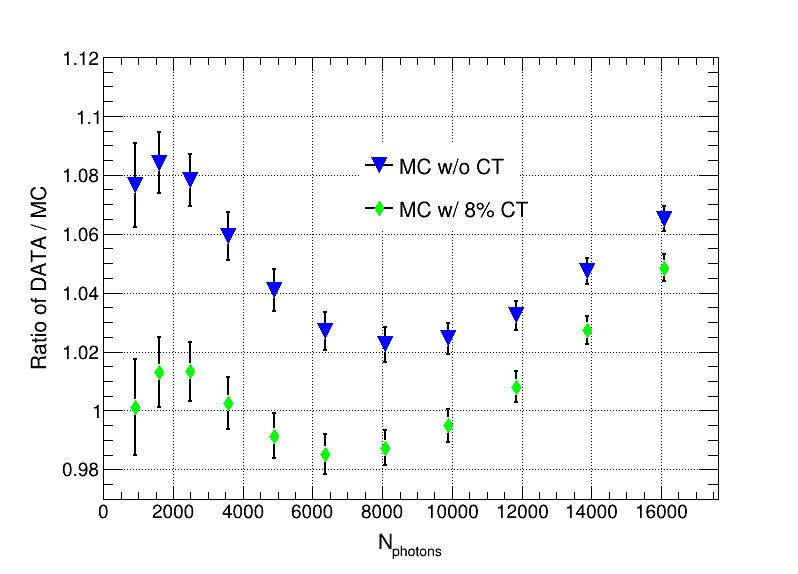}

\caption{\label{fig:measurement} \textit{Left}: Schematic of LED calibration system.  \textit{Middle}: SiPM (with 10,000 pixels) response curve for digital model results and real test data. \textit{Right}: Ratio of SiPM response by digital model and real test data. }
\end{figure}

An LED calibration system, as shown on the left-hand side of Figure~\ref{fig:measurement}, is designed to measure the real SiPM response.
The blue LED is driven by the FG3252 signal generator and provides a light pulse that is around 460 nm shorter than the 10-ns light pulse into the integrating sphere. Two beams of light are emitted; one is incident on the surface of the SiPMs under testing, and the other is monitored with the PMT calibration system.
The PMT calibration system can achieve a dynamic range of more than 0.5 million for the readout through three different dynodes. When the number of incident SiPM surface photons is low (one-tenth of the total number of pixels), the output of the SiPM is nearly linear and is used to calibrate the PMT calibration system. When the number of incident SiPM surface photons is high, the SiPM produces a saturation response; nevertheless, the PMT calibration system maintains good linearity. Therefore, the output of the PMT calibration system can monitor the number of incident SiPM surface photons.
The response of the SiPM under testing can also be obtained by the digitization program with the measured cross-talk probability in the above-mentioned step and known number of pixels.
The real SiPM response measurement result and MC results are shown in Figure~\ref{fig:measurement} (middle).
The right-hand side of Figure~\ref{fig:measurement} shows the ratio of the MC data as a function of the number of incident photons. By comparing the results with and without cross-talk, the cross-talk probability is found to affect the digitization results by a small percentage.
With the correct cross-talk probability, the digitization results agree with the actual SiPM response within 2\% in the range of 0--12,000 photons for 10,000 SiPM pixels.
The intensity of the LED light pulse increases with the voltage and the width increases at the same time, which may lead to a second-time avalanche at the same pixel.
Therefore, the actual SiPM response exceeds the MC response when the number of incident photons exceeds 12,000.

\subsection{Correction of SiPM saturation response}
\label{subsec:feature}
\paragraph{} Based on the SiPM response curve equation, i.e., Eq. \eqref{eq:SiPMformula}, the saturation response can be corrected to achieve linearity.
The number of incident photons $N_{photon}$ is calculated from the number of fired pixels $N_{fired}$ according to the following equation:

\begin{equation}
\label{eq:correction}
\begin{split}
N_{photon} &= N_{eff}\cdot[\ln(N_{eff})-\ln(N_{eff}-N_{fired})]
\end{split}
\end{equation}

The parameter $N_{eff}$ is obtained by fitting the real SiPM response curve; therefore, this parameter determines the accuracy of the correction.
As described above, $N_{eff}$ is a function of the number of pixels and cross-talk probability within one recovery period.
Based on the digitization program, the SiPM response is obtained under the determined settings by sampling 10,000 events and corrected event by event.
The uncertainty in the correction is calculated according to the width of the number of fired pixels. The uncertainty after the correction increases intuitively with the number of incident photons.

\begin{figure}[htbp]
\centering 
\includegraphics[width=.45\textwidth]{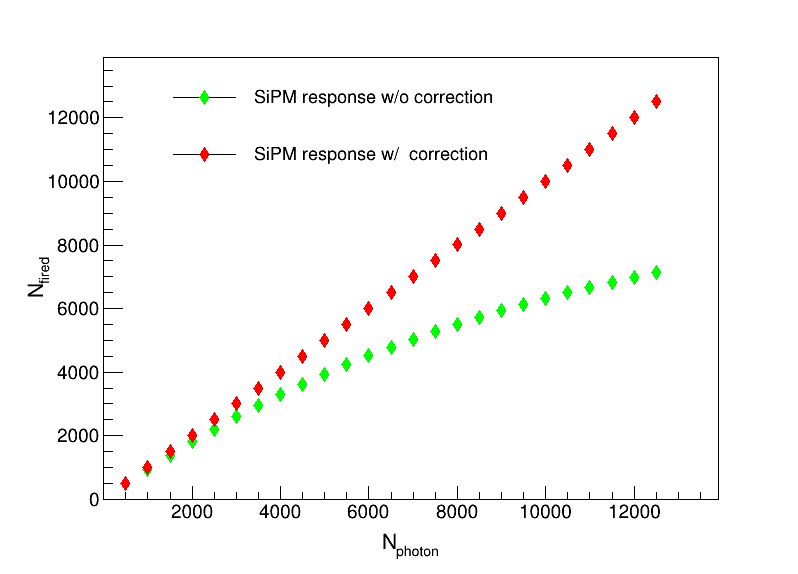}
\qquad
\includegraphics[width=.45\textwidth]{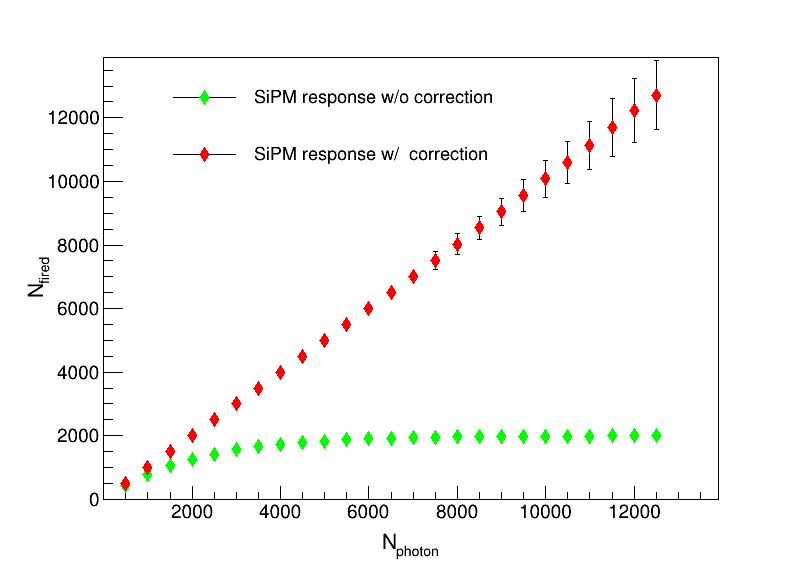}
\caption{\label{fig:SiPMmodel} SiPM response curve before correction (green boxes) and after correction (red boxes) for 10,000 pixels (left) and 2,000 pixels (right). Each point has 10,000 sampling events. The uncertainty is calculated according to the width of the $N_{fired}$ distribution. The relative width for 10,000 pixels is less than $1 \%$ in the range 0--12,000 photons.}
\end{figure}

Two examples of the SiPM response by MC simulation are shown in Figure~\ref{fig:SiPMmodel}.
These results indicate that although the saturation of SiPMs can be corrected accurately, a smaller number of pixels induce much uncertainty after correction even though the mean value is maintained. The relative width is less than $1 \%$ when the effective number of photons is less than the number of pixels, and it reaches around $10 \%$ when the effective number of photons is around six times the number of pixels.
When the SiPMs are exposed to a large number of incident photons, all the pixels may fire, resulting in SiPM oversaturation that cannot be corrected.
Therefore, the correction has an upper limit in terms of the number of incident photons, i.e., around six times the number of pixels.

\section{Full Simulation of Higgs measurement}
\label{sec:simu}

\subsection{ Higgs boson mass reconstruction and resolution}

\begin{figure}[htbp]
\centering 
\includegraphics[width=.45\textwidth]{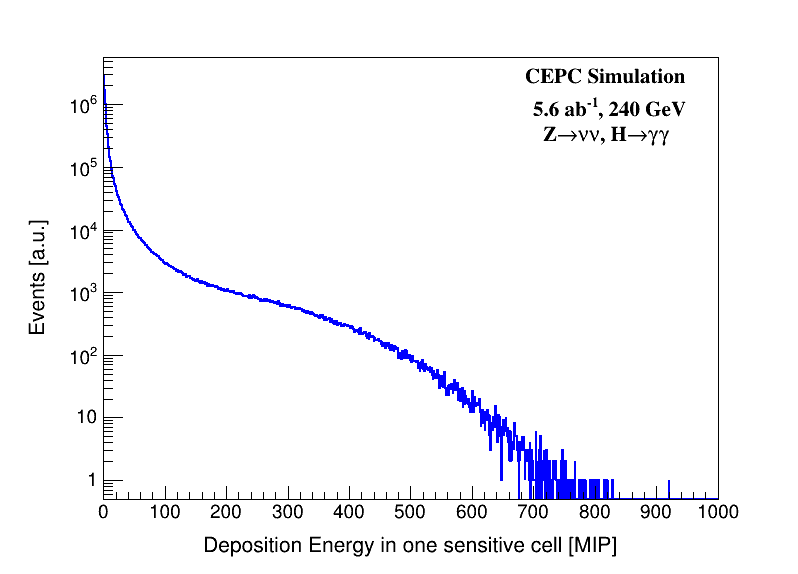}
\qquad
\includegraphics[width=.45\textwidth]{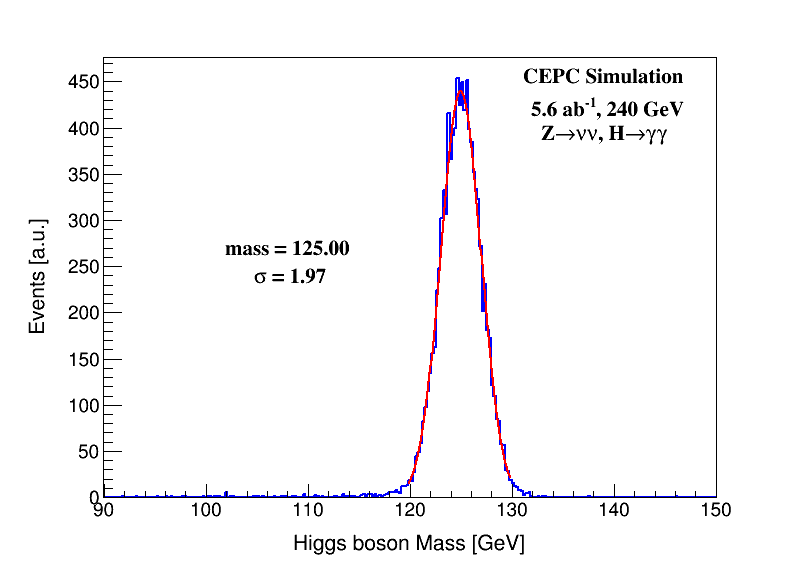}
\caption{\label{fig:higgsMass} \textit{Left}: deposition energy of H$\rightarrow$$\gamma$$\gamma$ in one sensitive cell in MIP unit; the required dynamic range is 1--800 MIP. \textit{Right}: The Higgs boson mass reconstructed without the effect of SiPM saturation. 10,000 Z$\rightarrow$$\nu$$\nu$,H$\rightarrow$$\gamma$$\gamma$ are generated in the simulation.}
\end{figure}

\paragraph{}The CEPC as a Higgs factory is designed to operate at a center-of-mass energy of $240 GeV$ and deliver an integrated luminosity of $5.6\  ab^{-1}$ in 7 years.
The $e^+e^-$ $\rightarrow$ $ZH$ process Z$\rightarrow$$\nu$$\nu$, H$\rightarrow$$\gamma$$\gamma$ channel only involves electromagnetic objects and strongly depends on the performance of the electromagnetic calorimeter. This channel is an ideal physics process that is considered as an electromagnetic calorimeter benchmark.  The Arbor algorithm~\cite{5} is adopted to reconstruct the electromagnetic showers of two photons under the CEPC software framework.
The deposition energy of the final two photons in one sensitive cell in ScECAL is shown on the left-hand side of Figure~\ref{fig:higgsMass}; the dynamic range was 800 MIPs. Therefore, ScECAL must have good linearity for around three orders of magnitudes for final-state electromagnetic objects.
The width of the reconstructed $H$$\rightarrow$$\gamma$$\gamma$ invariant mass with ScECAL is 1.97 GeV corresponding to a relative resolution of 1.58\%, as shown on the right-hand side of Figure~\ref{fig:higgsMass}.

\begin{figure}[htbp]
\centering 
\includegraphics[width=.45\textwidth]{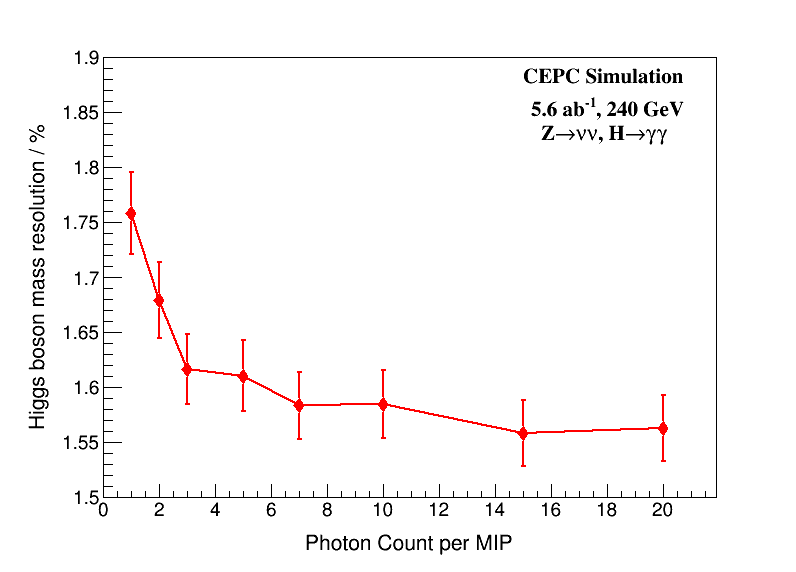}
\qquad
\includegraphics[width=.45\textwidth]{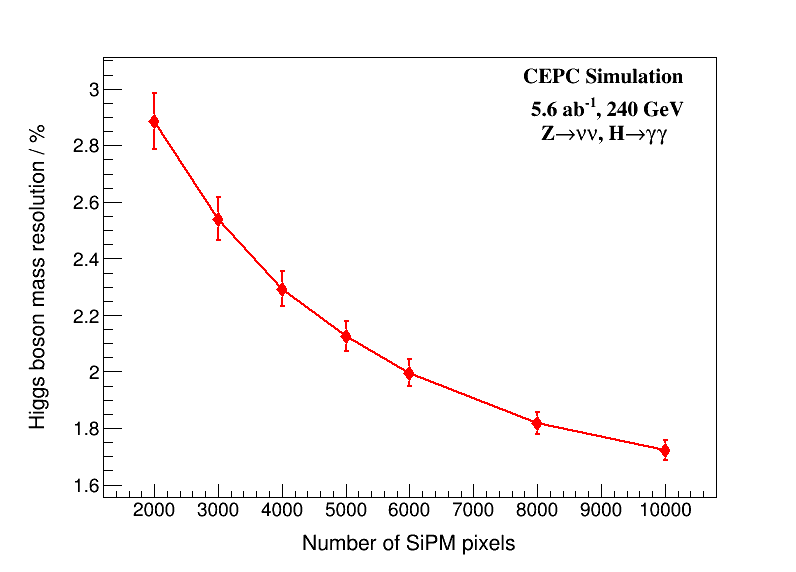}
\caption{\label{fig:lightyield} \textit{Left}: Relative resolution of Higgs boson mass with di-photon final state as a function of the photon count per MIP; the light yield should be no less than 10 p.e. per MIP in order to eliminate statistic fluctuation. \textit{Right}: Relative resolution of Higgs boson mass with di-photon final state as a function of the number of SiPM pixels without correction (20 p.e. per MIP); SiPM saturation has a non-trivial effect on the dynamic range.}
\end{figure}

Based on the CEPC simulation and reconstruction tools, the digitization process is implemented to study the effect of the SiPM response on the physics benchmark.
We know that the SiPM response is a function of the number of incident photons and pixels.
The number of incident photons is determined by the light yield, and the light yield per minimum ionization particles (MIP) is generally used as a metric in practice.
According to the reflective film, degree of polishing, and photon detection efficiency of SiPMs, the light yield can be optimized within a reasonable range.
A very low light yield will introduce nontrivial statistic fluctuation, whereas a very high light yield will decrease the dynamic range of SiPMs.
The Higgs boson mass relative resolution as a function of the light yield is shown on the left-hand side of Figure ~\ref{fig:lightyield}, which suggests that the statistic fluctuation is negligible when the light yield is greater than 10 p.e. per MIP. The minimum light yield requirement for an actual detector is to distinguish electronic noise and MIP signals clearly. In general, $10$--$20$ p.e. per MIP is essential for MIP separation. The maximum light yield is determined by the required dynamic range of SiPMs.
The right-hand side of Figure ~\ref{fig:lightyield} shows the Higgs boson mass relative resolution as a function of the number of SiPM pixels without correction.
SiPM saturation has a nontrivial impact on the Higgs boson mass relative resolution. Although increasing the number of pixels can ameliorate the saturation, a dedicated correction is also essential.

\begin{figure}[htbp]
\centering 
\includegraphics[width=.45\textwidth]{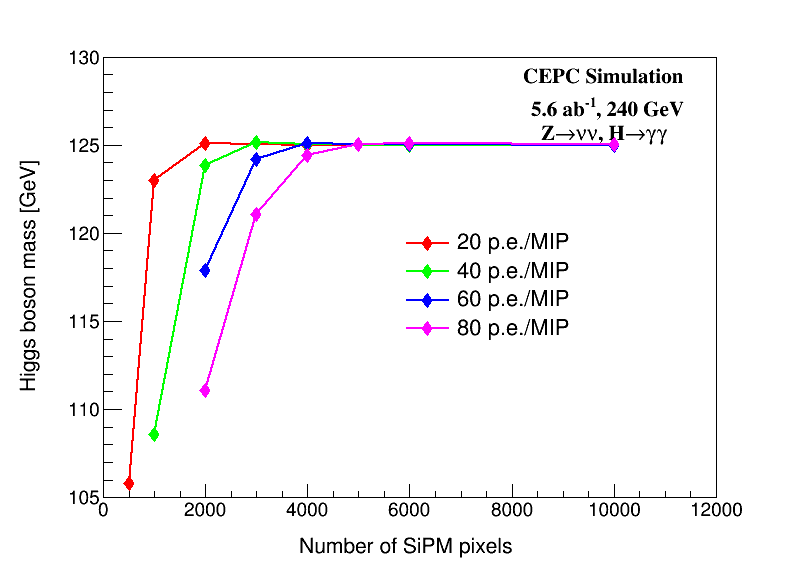}
\qquad
\includegraphics[width=.45\textwidth]{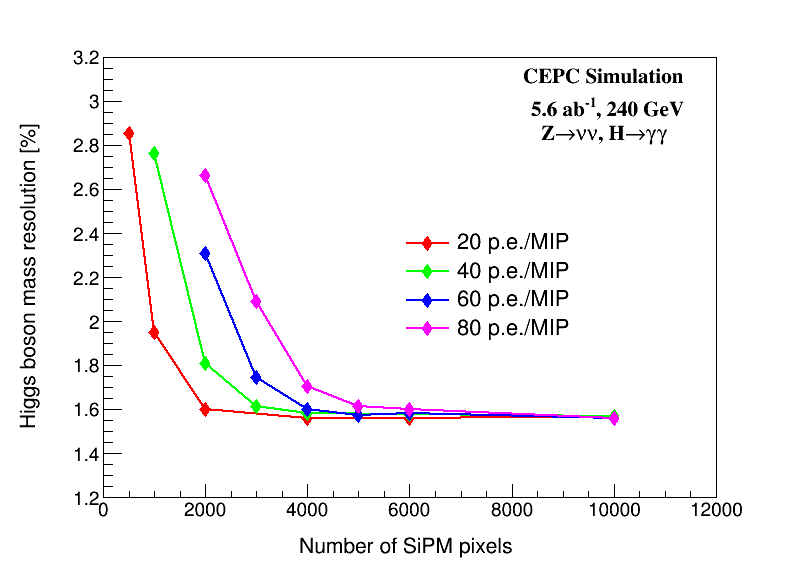}
\caption{\label{fig:HiggsMass} After correction of SiPM saturation response: \textit{Left}: Higgs boson mass as a function of the number of SiPM pixels. \textit{Right}: Relative resolution of Higgs boson mass with di-photon final state as a function of the number of SiPM pixels. The different colors represent the four different light yield scenarios.}
\end{figure}

Correction of the SiPM saturation response was implemented and examined in the simulation. Figure ~\ref{fig:HiggsMass} shows the Higgs boson mass and relative resolution as a function of the number of pixels in different light yield scenarios.
For each light yield scenario, the Higgs mass and relative resolution intuitively degraded with the decrease in the number of pixels and tended to a stable value when the number of pixels was sufficiently large.
This trend suggests that the correction of the SiPM saturation response is limited. When the number of pixels is very small, some of the sensitive cells with high deposition energy may cause SiPM oversaturation, which cannot be corrected to achieve good linearity.
Points are defined as the critical number of pixels when the Higgs boson mass and relative resolution begin to stabilize.
The critical number of pixels increases with the light yield and is expressed as follows: \eqref{eq:CriticalPixels}
\begin{equation}
\label{eq:CriticalPixels}
\begin{split}
N_{pixel} &= 50\cdot N_{LY}+1000
\end{split}
\end{equation}

For the CEPC ScECAL, the typical light yield in one sensitive cell is around 20 p.e. per MIP~\cite{12}.
With a light yield of 20 p.e. per MIP, we recommend that the number of SiPM pixels should be no less than 4000 in order to allow for some redundancy.


\subsection{Higgs measurement sensitivity from background}

\begin{figure}[htbp]
\centering
\includegraphics[width=.45\textwidth]{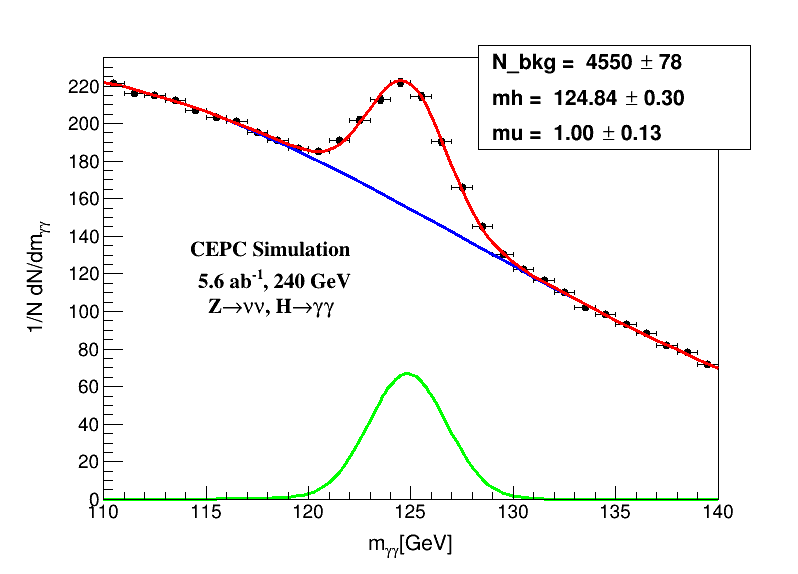}
\qquad
\includegraphics[width=.45\textwidth]{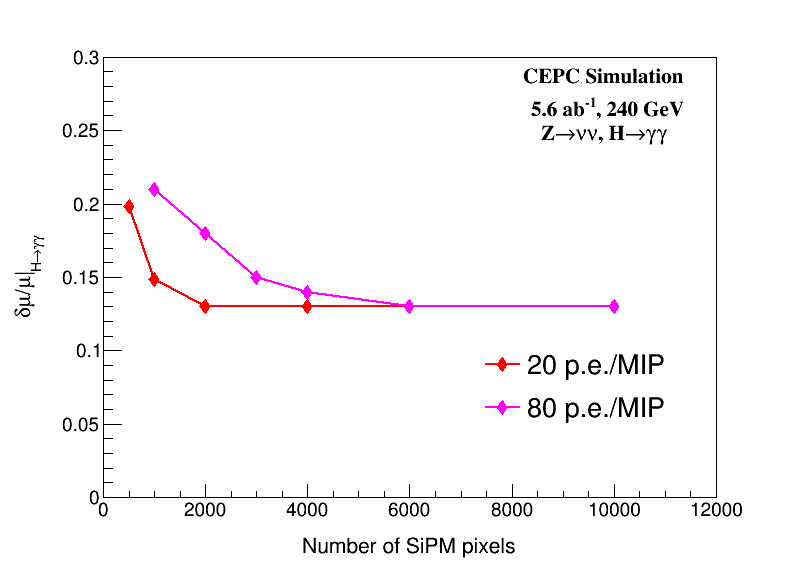}
\qquad
\caption{\label{fig:globalFit} \textit{Left}: $H \rightarrow$$\gamma$$\gamma$ resonance over a smooth background in the diphoton mass distribution. \textit{Right}: Sensitivity to $H \rightarrow$$\gamma$$\gamma$ versus the number of SiPM pixels. The markers and their uncertainties represent the expectations from a CEPC dataset of $5.6  ab^{-1}$  with 321.5 signal events selected.}
\end{figure}

\paragraph{}
Different decay modes of the Higgs boson can be identified through their unique signatures over the background from the Standard Model production~\cite{13}. The relative precision of the Higgs signature of the Standard Model expectation is defined as the measurement sensitivity.
The dominant background of the Z$\rightarrow$$\nu$$\nu$, H$\rightarrow$$\gamma$$\gamma$ channel from the $ZH$ process comprises single $Z$ production photons that arise from the initial and final state radiation. The background photons have an extensive energy range of $100 MeV$ to $120 GeV$. The effect of the SiPM response on the sensitivity to $H \rightarrow$$\gamma$$\gamma$ was also examined.
Resonance over a smooth background in the diphoton mass distribution is expected to appear for the $H \rightarrow$$\gamma$$\gamma$ decay channel, as shown in Figure \ref{fig:globalFit}.
A global fit is performed to estimate the sensitivity to $H \rightarrow$$\gamma$$\gamma$ measurement. The right-hand side of Figure \ref{fig:globalFit} shows the sensitivity as a function of the number of pixels in two different light yield scenarios. The critical number of pixels agrees with the Higgs boson mass measurement. No less than 4000 pixels are essential for CEPC ScECAL with a light yield of 20 p.e. per MIP.

\section{Summary and Discussion}
\label{sec:summary}

\paragraph{}
The rich physics at the Circular Electron Positron Collider (CEPC) requires a highly homogenous and stable electromagnetic calorimeter (ECAL) system for efficient separation of final-state particle showers and precise measurement of the induced energy with good linearity and resolution, especially for electromagnetic objects.
The PFA-oriented scintillator-based ScECAL is a promising candidate for the CEPC detector system.
Using SiPM technology, the ScECAL provides a high-granularity calorimeter that can efficiently separate nearby showers. Technologically mature and massively produced SiPM sensors are important for the stability and cost-effectiveness of the ScECAL.
Meanwhile, the SiPM sensors have a finite number of pixels, causing significant saturation behavior under high light exposure.
To precisely reconstruct the energy of the incident energetic particles at the ScECAL, it is essential to understand and correct this saturation effect of SiPM sensors.
We developed a digitization program for the SiPM response, which is a function of the number of incident photons and number of pixels.
This model predicts a significant saturation effect under high light exposure for SiPM design.
Compared with the experimental data, this model exhibits an agreement within 2\% for a 10,000-pixel SiPM in the range of 12,000 incident photons.
Base on this model, we developed a correction algorithm and calculated the uncertainty in the correction.
The correction has an upper limit in terms of the number of pixels, i.e., around six times the number of incident photons, as all the pixels may fire when the SiPM is exposed to a large number of photons, and correction will not be possible.
Using the CEPC full simulation software, we implemented this model and estimated the performance degradation for the physics benchmark of $ZH$, Z$\rightarrow$$\nu$$\nu$, H$\rightarrow$$\gamma$$\gamma$ at $240 GeV$ c.m.s.
Our results showed that the Higgs boson mass and relative resolution are a function of the number of SiPM pixels and light yield.
Points are defined as the critical number of pixels when the Higgs boson mass and relative resolution begin to stabilize.
The approximation formula $N_{pixel} = 50\cdot N_{LY}+1000$ was shown to describe the relationship between the critical number of pixels and the light yield per MIP unit.
For the CEPC ScECAL with a typical light yield of 20 p.e. per MIP, no less than 4000 pixels is recommended for precise Higgs measurement at the CEPC as well as to allow for redundancy.

This model is too idealistic; the reality is far more complex. The main implications of this study are as follows.
The formula of the SiPM response determines the measurement accuracy after correction and is obtained by measuring and fitting several SiPM response samples.
Therefore, feature homogeneity among different SiPMs, such as cross-talk probability, will degrade the measurement accuracy.
In addition, if the incident photons have a greater distribution than one recovery period, a second-time avalanche may occur in some pixels and affect the SiPM response curve.
Careful calibration is essential for correction of the SiPM saturation response in practice: SiPM gain calibration by measuring the single-photon spectrum from cell to cell, in-situ temperature monitoring, and SiPM gain correction.
A complete ScECAL technological prototype that includes 5040 10000-pixel SiPMs and 1260 4489-pixel SiPMs will be developed and demonstrated in practice.


\acknowledgments

\paragraph{Note added.} The author would like to thank Prof. Yanwen Liu of USTC for insightful discussions. The author also acknowledges the CEPC software group for providing essential simulation and analysis tools. This study was supported by the National Key Program for S\&T Research and Development (Grant No.: 2016YFA0400400) and National Natural Science Foundation of China (11635007).



\end{document}